\documentclass[twocolumn,showpacs]{revtex4}
\usepackage{graphicx}
\usepackage{amsmath}
\usepackage{amsfonts}
\usepackage{amssymb}
\usepackage{CJK}
\begin{document}
\title{Three types of statistics and the entropy bounds}
\author{Yong Xiao}
\email{xiaoyong@hbu.edu.cn}
\address{College of Physical Science and Technology, Hebei University, Baoding 071002, China}
\author{Yi-Xin Chen}
\address{Zhejiang Institute of Modern Physics, Zhejiang University, Hangzhou 310027, China}

\begin{abstract}
We investigated the entropy bounds of the three types of statistics:
para-Bose, para-Fermi and infinite statistics. We showed that the
entropy bounds of the conventional Bose, Fermi statistics and their
generalizations to parastatistics obey the $A^{3/4}$ law, while the
entropy bound of infinite statistics obeys the area law. This
suggests a close relationship between infinite statistics and
quantum gravity.
\end{abstract}
\maketitle

\section{Introduction\label{sec0}}
A systematic classification indicated that there are only three
types of consistent statistics in greater than two space dimensions:
para-Bose, para-Fermi and infinite statistics, with the former two
statistics as direct generalizations of Bose and Fermi statistics
\cite{haag,greenberg}. The algebras of Bose, Fermi and infinite
statistics can be viewed as the special cases of the $q$-deformed
commutation relation
$a_{k}a_{l}^{\dag}-qa_{l}^{\dag}a_{k}=\delta_{kl}$ with $q=1,-1$ and
$0$ \cite{greenberg}. While Bose and Fermi statistics are familiar
in the standard model of particle physics, infinite statistics has
becoming increasingly
attractive in recent years. Infinite statistics with $a_{k}a_{l}^{\dag}%
=\delta_{kl}$ has a great many interesting properties. Though there
is an obvious absence of locality in the theory of particles obeying
infinite statistics, other important properties like cluster
decomposition and the CPT theorem still hold which makes it capable
to be a sensible field theory. The nonlocality of the theory of
infinite statistics might be a virtue in the context of quantum
gravity for that it provides a new way in searching new physics
beyond local quantum field theory which is based on bosons and
fermions. At present, infinite statistics has been applied on the
discussions of black hole statistics \cite{stro,voloBH,Mbh,minicBH}
and dark energy quanta \cite{ng1,ng2,minicDE}.

It is known that bosonic and fermionic systems under the
gravitational stability condition are subject to the entropy bound
$A^{3/4}$ (we set $G,\hbar,c,k_{B}=1$ throughout), where $A$ is the
boundary area of the corresponding systems. The $A^{3/4}$ bound was
first derived from a photon gas system by 't Hooft \cite{hooft} and
has been verified in more general contexts
\cite{cohen,hsuEB,us1,us2,hsuMonster}. However, holographic
principle tells that the maximum entropy contained in a region
should be the area of its boundary in Planck units
\cite{hooft,susskind,bousso}. There seems to be an entropy gap
between the $A^{3/4}$ bound and the holographic entropy $A$. This
entropy gap even has its cosmological counterpart, which brings
about a huge numerical differences from $10^{90}$ to $10^{120}$ at
the present era of the universe \cite{barrow}. It is natural to ask
which mechanism can account for these absent degrees of freedom and
whether it is a part of a complete theory of quantum gravity.

In this paper, we concentrate on the entropy bounds attached to the
three types of statistics.  We shall derive the $A^{3/4}$ bound for
Bose, fermi statistics and their generalizations to parastatistics.
Then we turn to the case of infinite statistics and study its
entropy bound which was not addressed before. Our main result is
that the entropy bound of infinite statistics obeys the area law,
just as the holographic principle requires. We shall also discuss
the implications of this result.

\section{Bose, Fermi statistics and the $A^{3/4}$ bound \label{sec1}}

The entropy of bosonic and fermionic systems obeys the $A^{3/4}$
bound. This bound was first derived by 't Hooft by considering a
thermal photon gas confined to a box of size $l$. By statistics
mechanics, this system has entropy $S\sim l^{3}T^{3}$ and energy
$E\sim l^{3}T^{4}$. If there is no further limitation, one finds the
entropy of the system is proportional to the volume $l^{3}$ and
there is no bound to the entropy due to the arbitrary $T$. However,
according to general relativity, the energy of the system cannot
exceed the energy of a black hole of the same size. It gives $E\sim
l^{3}T^{4}\leqslant E_{bh}\sim l$ and leads to a critical
temperature $T\sim l^{-1/2}$. Substituting it into the entropy
formula, one easily finds the entropy bound $S_{\max}\sim A^{3/4}$,
where $A\sim l^{2}$ is the boundary area of the system.

For later comparison with the analysis of infinite statistics, we
give another derivation to the entropy bound $A^{3/4}$. It is known
that Bose and Fermi statistics in the high temperature limit can be
viewed as Boltzmann statistics except that the Gibbs factor
$\frac{1}{N!}$ is introduced. This factor is to offset the extra
degrees of freedom of Boltzmann statistics caused by particle
exchanging. Now we start from this equivalent statistics and discuss
the corresponding entropy bound. The canonical partition function
for a perfect gas of $N$ particles obeying this statistics is
\begin{align}
\begin{split}
Z_{N}&=\frac{1}{N!}\left( \sum_{i}^{\infty}e^{-\beta w_{i}}\right)
^{N}=\frac{1}{N!}\left( l^{3}\int e^{-\beta w}w^{2}dw\right)
^{N}\\
&\sim \frac{1}{N!}(l^{3}T^{3})^{N},\label{part}
\end{split}\end{align} where $T\equiv\beta^{-1}$. Note that we are
considering massless particles, for that we aim to derive an entropy
bound and the systems composed of particles with mass generally have
less entropy. The free energy is thus $F=-T\ln Z_{N}\sim-NT\ln\left(
\frac{l^{3}T^{3}}{N}\right) $. Its complete form is $-NT\left(
\ln(\frac{l^{3}T^{3}}{N})+1\right) $, but we omit all those
irrelevant coefficients to make the scaling behavior clear. Now we
get the energy and the entropy of the system
\begin{align}
E &  =-\left(  \frac{\partial\ln Z}{\partial\beta}\right) _{V,N}\sim
NT,\label{e1}\\
S &  =\ln Z + \beta E \sim N\ln\left(
\frac{l^{3}T^{3}}{N}\right).\label{s1}
\end{align}
From Eq.\eqref{e1} and Eq.\eqref{s1}, one can easily get
\begin{align}
S\sim N\ln\left( \frac{l^{3}E^{3}}{N^{4}}\right),
\end{align}
Fixing $E,l$ and varying the entropy with respect to $N$, we get the
maximum entropy $S\sim \left( El\right) ^{3/4}$. Taking into account
the non-gravitational collapse condition $E \leqslant E_{bh} \sim
l$, one obtains the critical temperature $T\sim l^{-1/2}$ and the
ultimate entropy bound
\begin{align}
S_{max}\sim\left( E_{bh}l\right) ^{3/4}\sim A^{3/4}.
\end{align}

It is worth to note one can also derive the entropy bound by
directly examining the dimension of the physically permitted Hilbert
space of the bosonic and fermionic fields \cite{us1}. For bosonic
fields, it means to count out the number of the field configurations
\begin{align}
\mid\Psi>=\mid n_{1},n_{2},n_{3}\cdots>,
\end{align}
which satisfies the non-gravitational collapse condition
\begin{align}
E_{\Psi}=\sum_{i=1}n_{i}w_{i}\leqslant E_{bh}\sim l. \label{g}
\end{align}
Here $n_{i}$ is the particle number occupying the mode of frequency
$w_i$. Due to the limitation \eqref{g}, the dimension of the Hilbert
space becomes finite and the realizable entropy finally has a bound
$A^{3/4}$ \cite{us1} . Moreover, the critical temperature $T \sim
l^{-1/2}$ also finds its explanation as an effective ultraviolet
cutoff $\Lambda$, which means the number of states with $w_{i}>
\Lambda$ being occupied is negligible compared to the number of
these states with only $w_{i}\leqslant \Lambda$ being occupied. See
\cite{us1} for details.

For completeness, here we turn to study the entropy bounds of
para-Bose and para-Fermi statistics. Parastatistics is characterized
by its order $p$ \cite{para1,para2}. For para-Bose statistics at
most $p$ particles can be in an antisymmetric state. For para-Fermi
statistics at most $p$ particles can be in a symmetric state. The
operator realization of parastatistics of order $p$ can be written
as
\begin{align}
a_{k}^{\dagger}=\sum_{\rho=1}^{p} b_{k}^{(\rho)\dagger},\  \
a_{k}=\sum _{\rho=1}^{p} b_{k}^{(\rho)}.
\end{align}
To describe parabosons (parafermions), $b_{k}^{(\rho)}$ and $b_{k}
^{(\sigma)\dagger}$ commute (anticommute) for $\rho=\sigma$ and
anticommute (commute) for $\rho\neq \sigma$. Obviously
parastatistics reduces to the common Bose and Fermi statistics in
the $p=1$ case.

Due to the statistical properties of para-Bose statistics, we write
the grand-canonical partition function for a collection of
para-bosons of order $p$ as
\begin{align}
\Xi=\prod_{i}\sum_{n_{i}=1}^{\infty} (1+p e^{-(\alpha+\beta
w_{i})n_{i}})=\prod_{i} \frac{1+(p-1)e^{-(\alpha+\beta
w_{i})}}{1-e^{-(\alpha+\beta w_{i})}}.
\end{align}
To derive the entropy bound, we consider massless particles and let
$\alpha=0$. Thus
\begin{align}
\begin{split}
\ln \Xi &\sim  l^{3} \int[\ln(1+(p-1)e^{-\beta w})  -
\ln(1-e^{-\beta w})]w^{2} dw\\
 &\sim c(p)l^{3} T^{3} .
\end{split}
\end{align}
The value of $c(p)$ is a positive number which can be approximately
evaluated and it monotonically increases with increasing $p$. The
energy and entropy is
\begin{align}%
\begin{split}
E  &  =-\left(  \frac{\partial\ln\Xi}{\partial\beta}\right)
_{V}\sim
c(p)l^{3} T^{4},\\
S  &  =\ln\Xi+ \beta E \sim c(p) l^{3} T^{3}.
\end{split}
\end{align}
So there is $S\sim c(p)^{1/4}(El)^{3/4} \leq c(p)^{1/4} A^{3/4}$.
The grand-canonical partition function of a collection of
para-fermions can be written as
\begin{align}
 \Xi=\prod_{i}(\sum_{n_{i}=0,1}e^{-(\alpha+\beta
w_{i})n_{i}})^p=\prod _{i}(  1+e^{-(\alpha+\beta w_{i})})^p.
\end{align}
This easily leads to an entropy bound $S\sim p^{1/4} A^{3/4}$.
Accordingly, the order $p$ only contributes to the coefficients of
the expressions of the entropy bounds and doesn't change their
scaling behavior. The entropy bounds of parastatistics always has a
$A^{3/4}$ scaling.

All above confirms the validity of the $A^{3/4}$ bound. The local
quantum field theory describing bosons and fermions can only account
for a very small part of the holographic degrees of freedom. The
entropy gap between $A^{3/4}$ and $A$ implies the complete theory of
quantum gravity should include more physical elements beyond the
conventional local quantum field theory.

\section{Infinite statistics and the $A$ bound \label{sec1}}
We begin with a concise review of the elementary ingredients of
infinite statistics. The basic commutation relation of infinite
statistics is
\begin{align}
a_{k}a_{l}^\dag=\delta_{kl}.\label{algebra} \end{align} Assume the
existence of the vacuum state $|0>$ annihilated by all the
annihilators. The entire state space can be constructed by creation
operators acting on the vacuum state in sequence. A general $N$
particle state can be written as
\begin{align}
a_{j_{N}}^{\dag}\cdots a_{j_{1}}^{\dag}|0>.\label{basis}
\end{align}
The number operator $n_{i}$ has the form
\begin{align}
n_{i}= a_{i}^\dag a_{i}+\sum\limits_{k}a_{k}^\dag a_{i}^\dag
a_{i}a_{k} +\sum\limits_{k_{1},k_{2}}a_{k_{1}}^\dag a_{k_{2}}^\dag
a_{i}^\dag a_{i}a_{k_{2} }a_{k_{1}}+\cdots,
\end{align}
With the help of Eq.\eqref{algebra}, one can check easily
\begin{align}
[n_{i},a_{j}]=-\delta_{ij}a_{j}.\label{number}
\end{align}
The subtlety here is the existence of a recursion pattern:
$n_{i}=a_{i}^\dag a_{i}+\sum\limits_{k}a_{k}^\dag n_{i}a_{k}$. The
operator $n_i$ acting on a state gives the correct particle number
occupying the mode of frequency $w_{i}$. A summation gives the total
number operator and the total energy operator
\begin{align}
N = \sum \limits_{i}n_{i}, \ \  E  =\sum \limits_{i}n_{i}w_{i}.
\label{nnn}
\end{align}

For infinite statistics, particle exchanging can provide distinct
states. This property is very different from the conventional Bose
and Fermi statistics. It is observed from the orthogonal relation
\begin{align}
<0|a_{i_{1}}\cdots a_{i_{N}}a_{j_{N}}^\dag\cdots a_{j_{1}}^\dag
|0>=\delta_{i_{1}j_{1}}\cdots\delta_{i_{N}j_{N}},
\end{align}
of two $N$ particle states. This is an immediate result of the basic
commutation relation \eqref{algebra}. Since changing the order of
the particles gives another state orthogonal to the original one,
the particles obeying infinite statistics are virtually
distinguishable. This kind of distinguishability implies a
rediscovery of Boltzmann statistics. Hence, infinite statistics is
also called ``quantum Boltzmann statistics'' where ``quantum'' means
the phase space is quantized according to quantum mechanics. In
addition, infinite statistics can also be viewed as the statistical
property of identical particles with an infinite number of internal
degrees of freedom. The particles are distinguishable by their
internal states \cite{greenberg}.

Now we focus on the derivation of the entropy bound. Consider a
system of $N$ noninteracting massless particles obeying infinite
statistics. Compared with Eq.\eqref{part}, the Gibbs factor
$\frac{1}{N!}$ must be absent here for particle exchanging can
provide distinct quantum states in the theory of infinite
statistics. Thus the canonical partition function is written as
\begin{align}
\begin{split}
Z_{N}&=\left( \sum_{i}^{\infty}e^{-\beta w_{i}}\right) ^{N}=\left(
l^{3}\int e^{-\beta w}w^{2}dw\right)
^{N}\\
&\sim(l^{3}T^{3})^{N},
\end{split}
\end{align}
which leads to
\begin{align}
 E\sim NT,\   \   S\sim N\ln\left( l^{3}T^{3}\right).\label{infiS}
\end{align}
So we have
\begin{align}
S\sim N\ln\left( \frac{l^{3}E^{3} }{N^{3}}\right) ,
\end{align}
Fixing $E,l$ and varying the entropy with respect to $N$, we find
the maximum $S \sim El$. Interestingly, it takes the form as the
famous ``Bekenstein entropy bound'' \cite{bekenB1,bekenB2}. Imposing
the energy limitation from general relativity: $E\leqslant E_{bh}$,
we find that the critical temperature is at $T\sim l^{-1}$ and the
final entropy bound of infinite statistics obeys the area law
\begin{align}
S_{max} \sim E_{bh}l \sim A. \label{areaB}
\end{align}
It is an interesting result. At first sight, since exchanging
particles can provide new states, it is natural to expect infinite
statistics has a higher entropy bound than $A^{3/4}$, but how can
one expect it obeys the area law just as holographic principle
requires? Considering that infinite statistics is the only
consistent statistics other than Bose and fermi statistics, the
coincidence with the holographic principle is impressive.
Furthermore, here the critical temperature $T$ is around $l^{-1}$.
It means the behavior of the system is dominated by its long-wave
components. All the information inside the size $l$ of the system
can be smeared out, which is similar to that of black hole physics.
This partly explains why infinite statistics can be effectively used
to describe the properties of black holes as suggested in
\cite{Mbh,minicBH}.

As a supplement, here we make a verification to the result
\eqref{areaB} by direct counting the number of the microscopic
states contributing to the entropy. When massless fields are
confined to a region of size $l$, the momenta will be quantized as
$\vec{k}\sim \frac{1}{l}\left( m_{x},m_{y},m_{z}\right) $. The
frequencies of the modes are $w_{i}=|\vec{k_{i}}|$. Then we can use
$a_{k_i}^{\dag}$ to construct all the basis of the Hilbert space of
the infinite statistics fields. The general field configuration is
exhibited in Eq.\eqref{basis}. We impose a further limitation
$\sum\limits_{i=1}n_{i}w_{i}\leqslant E_{bh}$ to exclude the states
with energy larger than black hole for the Hilblert space to be
physically accessible. This requirement causes the dimension of the
Hilbert space to be finite.

The critical temperature $T\sim l^{-1}$ also serves as an effective
cutoff $\Lambda$, which means the dominate states to the entropy are
these with only $w_{i}\leqslant \Lambda$ being occupied. Hence we
can consider a simplified system which contains only the three
lowest momentum modes $\vec{k}_{1} \sim \frac{1 }{l}\left(
1,0,0\right) $, $\vec{k}_{2}\sim \frac{1}{l}\left( 0,1,0\right) $,
$\vec{k}_{3}\sim \frac{1}{l}\left( 0,0,1\right)  $. Now we can write
out all the corresponding field configurations as the form similar
to
\begin{align}
\overbrace{a_{k_{3}}^{\dag }a_{k_{2}}^{\dag }a_{k_{1}}^{\dag }\cdots
a_{k_{3}}^{\dag }a_{k_{2}}^{\dag }a_{k_{1}}^{\dag }}^{N}|0>,
\end{align}
The next step is to count the number of them. When the system has
the critical energy $E_{bh}$, the particle number $N$ of the system
is $N\sim \frac{E_{bh}}{1/l}\sim E_{bh}l\sim A.$ Sine the three
modes have the same energy, the most probable distribution of the
particles on these modes is each mode being occupied by
$\frac{N}{3}$ particles. Due to the exchanging property of infinite
statistics, the number of different field configurations is
\begin{align}
W=\frac{N!}{\frac{N}{3}!\frac{N}{3}!\frac{N}{3}!}\sim
\frac{N^{N}}{\left( \frac{N}{3}\right)  ^{N}}\sim3^{N}\sim3^{A},
\end{align}
Then we get an area entropy
\begin{align}
S=\ln W\sim E_{bh}l\sim A.
\end{align}
It supports our earlier result from thermodynamical analysis.

In fact, by choosing proper parameters it is not hard to obtain an
area-form entropy from infinite statistics. This fact is usually
exploited to discuss the entropy of black hole or dark energy
\cite{minicBH,ng1,ng2}. However, the bound to the entropy is another
meaningful question worthy to be addressed \cite{Mbh}. In some
practical context the entropy may appear to be unbounded. For
example, if one starts from a non-relativistic partition function
$Z_N =(l^{3}\left( mT\right) ^{3/2})^{N}$ for particles with mass,
the derived entropy is $S_{max}\sim mEl^2$. Though choosing particle
mass and energy to be $m\sim l^{-1}, E\sim l$ can lead to an area
entropy $S\sim A$, generally the entropy is unbounded. The key to
solve this problem is to consider the complete relativistic
partition function $Z_{N}\sim(l^{3}\left( mT\right)^{3/2}
e^{-m/T})^{N}$, one can check roughly the maximum entropy is
$S_{max}\sim\frac{ml}{1+m^{2}l^{2} }\left( El\right) <A$. It is
crucial to introduce the static mass term $e^{-m/T}$ to get the
bound.

\section{Conclusion remarks\label{sec4}}
We have examined the entropy bounds of the three types of
statistics. We showed that the entropy bounds of (para-)Bose and
(para-)Fermi statistics obey the $A^{3/4}$ law, while the bound of
infinite statistics obeys the area law. When a bosonic or fermionic
system collapses to form a black hole, the system will be controlled
by gravity and the entropy will evolve from $A^{3/4}$ to $A$. So the
entropy gap between $A^{3/4}$ to $A$ should be explained by a final
theory of quantum gravity. Now infinite statistics provides a new
way to fill up this gap. This suggests that there might be a
relationship between infinite statistics and quantum gravity.

The theory of infinite statistics has intriguing properties, such as
nonlocality and non-extensive entropy, which resemble these of
gravitational systems. So it is not weird to conjecture that
infinite statistics serves as an essential ingredient of the
complete theory of quantum gravity. For example, Strominger has
argued that the gas of charged extremal black holes should obey
infinite statistics \cite{stro}. In addition, the large $N$ limit of
$SU(N)$ matrix theory can be effectively described by the master
fields obeying the infinite statistics algebra $a_k
a_l^\dagger=\delta_{kl}$ \cite{voloBH,v1,v2,hal1,hal2,hal3}. Since
the large $N$ limit of $SU(N)$ theory is equivalent to the theory of
gravity by virtue of techniques like AdS/CFT, infinite statistics
should also play a part in quantum gravity. Whatever, at present
there are only a limited number of clues on the relationship between
infinite statistics and quantum gravity. Many open questions should
be further clarified.

By symmetrizing and anti-symmetrizing the state space of infinite
statistics, the bosonic and fermionic subspaces can be retrieved. So
there should be a transition mechanism from the infinite statistics
theory to the conventional quantum field theory describing bosons
and fermions. We notice there has been an attempt on this question,
which suggested infinite statistics is related to the new physics in
high energy scale and discussed the hierarchy problem from the
electroweak scale to the Planck scale \cite{shev}. This is also a
question worthy of further study.

\section*{Acknowledgements}
We would like to thank C. Cao, J.L. Li and Y.Q. Wang for useful
discussions. The work is supported in part by the NNSF of China
Grant No.11047120, No.11075138 and the Science Foundation of Hebei
University  No.2010Q29.

\end{document}